\documentclass[letterpaper, 12 pt]{article}

\usepackage[letterpaper,margin=0.75in]{geometry}

\usepackage{titlesec}
\titleformat*{\subsection}{\bf\itshape}

\usepackage{amssymb,amsmath,amstext,amsfonts}
\usepackage{mathtools}
\usepackage{bm}  
\usepackage{siunitx}
\usepackage{physics}

\usepackage{filecontents}
\usepackage{stmaryrd}

\usepackage{lettrine}

\usepackage{cite}
\usepackage{graphicx}
\usepackage{tikz}
\usepackage{textcomp}
\usepackage{xcolor}
\def\BibTeX{{\rm B\kern-.05em{\sc i\kern-.025em b}\kern-.08em
		T\kern-.1667em\lower.7ex\hbox{E}\kern-.125emX}}

\usepackage{multicol}
\usepackage{multirow}

\usepackage{url}

\usepackage{booktabs}
\usepackage{tabu}

\usepackage{environ}

\usepackage{enumerate}
\usepackage{ifthen}

\usepackage{xargs} 
\usepackage[colorinlistoftodos,prependcaption,textsize=tiny]{todonotes}
\newcommandx{\unsure}[2][1=]{\todo[linecolor=red,backgroundcolor=red!25,bordercolor=red,#1]{#2}}
\newcommandx{\change}[2][1=]{\todo[linecolor=blue,backgroundcolor=blue!25,bordercolor=blue,#1]{#2}}
\newcommandx{\info}[2][1=]{\todo[linecolor=OliveGreen,backgroundcolor=OliveGreen!25,bordercolor=OliveGreen,#1]{#2}}
\newcommandx{\improvement}[2][1=]{\todo[linecolor=Plum,backgroundcolor=Plum!25,bordercolor=Plum,#1]{#2}}
\newcommandx{\thiswillnotshow}[2][1=]{\todo[disable,#1]{#2}}

\usepackage{xargs} 
\newcommandx{\rotmatvar}[5][2=, 3=, 4=, 5=, usedefault=&]%
{{\ensuremath{{}_{#2}^{#3}{\bm{#1}}^{#4}_{#5}}}}
\newcommandx{\rotmatvardot}[5][2=, 3=, 4=, 5=, usedefault=&]%
{{\ensuremath{{}_{#2}^{#3}{\dot{\bm{#1}}}^{#4}_{#5}}}}
\newcommandx{\rotmatvarddot}[5][2=, 3=, 4=, 5=, usedefault=&]%
{{\ensuremath{{}_{#2}^{#3}{\ddot{\bm{#1}}}^{#4}_{#5}}}}
\newcommandx{\rotmatvarhat}[5][2=, 3=, 4=, 5=, usedefault=&]%
{{\ensuremath{{}_{#2}^{#3}{\hat{\bm{#1}}}^{#4}_{#5}}}}
\newcommandx{\rotmatvarbar}[5][2=, 3=, 4=, 5=, usedefault=&]%
{{\ensuremath{{}_{#2}^{#3}{\overline{\bm{#1}}}^{#4}_{#5}}}}
\newcommandx{\rotmatvartilde}[5][2=, 3=, 4=, 5=, usedefault=&]%
{{\ensuremath{{}_{#2}^{#3}{\tilde{\bm{#1}}}^{#4}_{#5}}}}
\newcommandx{\rotvecvar}[5][2=, 3=, 4=, 5=, usedefault=&]%
{{\ensuremath{{}_{#2}^{#3}{\bm{#1}}^{#4}_{#5}}}}
\newcommandx{\rotsymbvar}[5][2=, 3=, 4=, 5=, usedefault=&]%
{{\ensuremath{{}_{#2}^{#3}{\bm{#1}}^{#4}_{#5}}}}
\newcommandx{\elemHomVar}[3][1=Rot, 2=z, 3=\ang{90}, usedefault=&]%
{{\ensuremath{{\text{#1}}_{\text{#2}}(#3)}}}

\newcommandx{\elemRotZ}[1][1=\varphi, usedefault= & ]%
{{\ensuremath{%
      \begin{bmatrix}
        c_{#1} & -s_{#1} & 0 \\
        s_{#1} & c_{#1}  & 0 \\
        0      & 0       & 1
      \end{bmatrix}%
    }}}
\newcommandx{\elemRotY}[1][1=\vartheta, usedefault= & ]%
{{\ensuremath{%
      \begin{bmatrix}
        c_{#1}  & 0 & s_{#1} \\
        0       & 1 & 0      \\
        -s_{#1} & 0 & c_{#1}
      \end{bmatrix}%
    }}}
\newcommandx{\elemRotX}[1][1=\psi, usedefault= & ]%
{{\ensuremath{%
      \begin{bmatrix}
        1 & 0      & 0       \\
        0 & c_{#1} & -s_{#1} \\
        0 & s_{#1} & c_{#1}   
      \end{bmatrix}%
    }}}

\newcommandx{\elemHomRotZ}[1][1=\varphi, usedefault= & ]%
{{\ensuremath{%
      \begin{bmatrix}
        c_{#1} & -s_{#1} & 0 & 0 \\
        s_{#1} & c_{#1}  & 0 & 0 \\
        0      & 0       & 1 & 0 \\
        0      & 0       & 0 & 1
      \end{bmatrix}%
    }}}
\newcommandx{\elemHomRotY}[1][1=\vartheta, usedefault= & ]%
{{\ensuremath{%
      \begin{bmatrix}
        c_{#1}  & 0 & s_{#1} & 0 \\
        0       & 1 & 0      & 0 \\
        -s_{#1} & 0 & c_{#1} & 0 \\
        0       & 0 & 0      & 1
      \end{bmatrix}%
    }}}
\newcommandx{\elemHomRotX}[1][1=\psi, usedefault= & ]%
{{\ensuremath{%
      \begin{bmatrix}
        1 & 0      & 0       & 0 \\
        0 & c_{#1} & -s_{#1} & 0 \\
        0 & s_{#1} & c_{#1}  & 0 \\
        0 & 0      & 0       & 1 
      \end{bmatrix}%
    }}}
\newcommandx{\elemHomTransZ}[1][1=c, usedefault= & ]%
{{\ensuremath{%
      \begin{bmatrix}
        1 & 0 & 0 & 0    \\
        0 & 1 & 0 & 0    \\
        0 & 0 & 1 & {#1} \\
        0 & 0 & 0 & 1
      \end{bmatrix}%
    }}}
\newcommandx{\elemHomTransY}[1][1=b, usedefault= & ]%
{{\ensuremath{%
      \begin{bmatrix}
        1 & 0 & 0 & 0    \\
        0 & 1 & 0 & {#1} \\
        0 & 0 & 1 & 0    \\
        0 & 0 & 0 & 1
      \end{bmatrix}%
    }}}
\newcommandx{\elemHomTransX}[1][1=a, usedefault= & ]%
{{\ensuremath{%
      \begin{bmatrix}
        1 & 0 & 0 & {#1} \\
        0 & 1 & 0 & 0    \\
        0 & 0 & 1 & 0    \\
        0 & 0 & 0 & 1
      \end{bmatrix}%
    }}}

\newsavebox{\myparbox}
\newlength{\myparboxwidth}

\providecommand{\norm}[1]{\lVert#1\rVert}

\newcommand\Angle[1]{\setbox0=\hbox{$\mskip 7mu minus 4mu#1$}%
  \raise.21ex\hbox{$/$}\hskip-0.95ex\underline{\raise\dp0\hbox{\box0}}}

\ExplSyntaxOn

\NewDocumentCommand \vect { s o m }
 {
  \IfBooleanTF {#1}
   { \vectaux*{#3} }
   { \IfValueTF {#2} { \vectaux[#2]{#3} } { \vectaux{#3} } }
 }

\DeclarePairedDelimiterX \vectaux [1] {\lbrack} {\rbrack}
 { \, \dbacc_vect:n { #1 } \, }

\cs_new_protected:Npn \dbacc_vect:n #1
 {
  \seq_set_split:Nnn \l_tmpa_seq { , } { #1 }
  \seq_use:Nn \l_tmpa_seq { \enspace }
 }
\ExplSyntaxOff

\usepackage[firstpageonly=true]{draftwatermark}
\usepackage{hyperref}
\hypersetup{
    colorlinks=true,
    linkcolor=blue,
    filecolor=magenta,      
    urlcolor=cyan,
    bookmarks=true,
}
\usepackage[noabbrev]{cleveref}  

\usepackage{algorithm,algpseudocode,float}
\usepackage{setspace} 

\algnewcommand{\algorithmicand}{\textbf{ and }}
\algnewcommand{\algorithmicor}{\textbf{ or }}
\algnewcommand{\AlgAnd}{\algorithmicand}
\algnewcommand{\AlgOr}{\algorithmicor}

\graphicspath{%
  {Figures/}
  {Figures/TiKz/}
  {Figures/Ipe/}
  {Figures/XFig/}
  {Figures/Inkscape/}
  {Figures/Matlab/}
  {Figures/Scanned/}
  {Figures/Saved/}
}

\DeclareMathOperator*{\argmin}{arg\,min}

\newcommand\footnotewithoutmark[1]{%
  \begingroup
  \renewcommand\thefootnote{}\footnote{#1}%
  \addtocounter{footnote}{-1}%
  \endgroup
  }

\title{Responding to Illegal Activities Along the Canadian Coastlines Using Reinforcement Learning}

\author{Mohammed Abouheaf, Shuzheng Qu, Wail Gueaieb,\\ Rami Abielmona, and Moufid Harb
\footnotewithoutmark{This work was partially supported by NSERC Grant~EGP~537568-2018.}%
\footnotewithoutmark{Mohammed Abouheaf, Shuzheng Qu and Wail Gueaieb are with the School of Electrical Engineering \& Computer Science, University of Ottawa, Ottawa, Canada. E-mail:~\{mabouhea,fqu096,wgueaieb\}@uottawa.ca.
Rami Abielmona and Moufid Harb are with Larus Technologies, Ottawa, Ontario, Canada. E-mail:\{rami.abielmona,moufid.harb\}@larus.com. 
}%
}
\date{}

\begin{document}
\maketitle
\thispagestyle{empty}
\pagestyle{empty}

\DraftwatermarkOptions{%
angle=0,
hpos=0.5\paperwidth,
vpos=0.97\paperheight,
fontsize=0.012\paperwidth,
color={[gray]{0.2}},
text={
  \newcommand{\thispaperdoi}{10.1109/MIM.2021.9400967}
  \newcommand{\thispaperCopyrightYear}{2021}
  \parbox{0.99\textwidth}{This is the postscript version of the published paper. (doi: \href{http://dx.doi.org/\thispaperdoi}{\thispaperdoi})\\
    \copyright~\thispaperCopyrightYear~IEEE.  Personal use of this material is permitted.  Permission from IEEE must be obtained for all other uses, in any current or future media, including reprinting/republishing this material for advertising or promotional purposes, creating new collective works, for resale or redistribution to servers or lists, or reuse of any copyrighted component of this work in other works.}},
}

Machine learning (ML) algorithms can prove to be instrumental in certain complex ill-conditioned systems when inserted as a middle layer to interface low-level hardware, such as sensors and actuators, and high-level decision-making kernels. Such an interface provides a secondary, or supervisory, conditioning layer that would enhance the system's robustness in the face of various types of uncertainties and disturbances. This article elaborates how ML can leverage the solution of a contemporary problem related to the security of maritime domains.


The worldwide ``Illegal, Unreported, and Unregulated'' (IUU) fishing incidents have led to serious environmental and economic consequences which involve drastic changes in our ecosystems in addition to financial losses caused by the depletion of natural resources. The term ``illegal'' refers either to contravening the fishing regulations or conducting fishing activities unlawfully under the jurisdiction of a certain maritime territory. In this context, ``unreported'' indicates the failure to report to the relevant authorities, while ``unregulated'' describes fishing activities in areas with no applicable measures to control the catch. The Fisheries and Aquatic Department (FAD) of the United Nation's Food and Agriculture Organization (FAO) issued a report which indicated that the annual losses due to IUU fishing reached $\$25$~Billion~\cite{FAO2016}. This imposes negative impacts on the future-biodiversity of the marine ecosystem and domestic Gross National Product (GNP). Maritime Domain Awareness (MDA) is concerned with the situational awareness relevant to the activities that have direct or indirect impacts on the operational, organizational, economical, and safety sides of the maritime domains~\cite{Rami2013,IMSANC}. The assets in MDA support different missions, such as monitoring, risk assessment, and management. Hence, robust interception mechanisms are increasingly needed for detecting and pursuing the unrelenting illegal fishing incidents in maritime territories.

\section*{Decision-Making System}
%
The IUU fishing problem can be broken down into at least three sub-problems:
\begin{enumerate}
\item Recognizing an IUU activity, where radar and satellite monitoring systems are deployed to detect the unreported event and provide live information about its status, such as the vessels' location measurements, for instance.
\item Deciding on the resources (coast guard vessels, military planes, etc.) to dispatch in the pursuit of the IUU vessels. The decision may depend on many factors, including the locations of the pursuing vehicles, their speeds, cost of operation, and fuel consumption, to name a few. Herein, autonomous sea vessels are considered to pursuit the illegal shipping activity.
\item Coordinating the motion of the fleet of pursuers to catch the IUU vessel while still in local waters.
\end{enumerate}

In this article, only the latter task is addressed. It is formulated as a pursuer-evader problem that is tackled within an ML framework. One or more pursuers, such as law enforcement vessels, intercept an evader (i.e., the illegal fishing ship) using an online reinforcement learning mechanism that is based on a value iteration process. It employs real-time navigation measurements of the evader ship as well as those of the pursuing vessels and returns back model-free interception strategies. 
%
Means of adaptive critics are employed to approximate the learning process using actor-critic neural networks, where a gradient descent approach is followed to tune the weights of the neural networks. Test cases are simulated with the aid of Total::Perception™ (T::P) engine, a professional predictive analytics decision-making software that is developed to mimic the situational awareness in maritime as well as other domains~\cite{Larus,Rami2017}.

%

\section*{Machine Learning Literature}


The significant advances in the architectures of digital processing units, enabled the implementation of sophisticated artificial intelligence algorithms on integrated microprocessors~\cite{IMFogel}. Such a technological breakthrough opened a new window of engineering application opportunities that were not possible a few years ago. This includes, for example, the visual detection and autonomous refueling of UAVs~\cite{IMSun}, robot motion planing~\cite{IMRLGomez}, brain MRI estimation methods~\cite{IMChap}, and latency measurement of distributed networks~\cite{IMMoh}, to name a few.
In a particular context, Approximate Dynamic Programming (ADP) is a machine learning tool that has been employed to tackle the computational challenges associated with the action and state spaces of the dynamic programming solutions. ADP can be classified into four classes, namely, Heuristic Dynamic Programming, Dual Heuristic Dynamic Programming, Action Dependent Heuristic Dynamic Programming, and Action Dependent Dual Heuristic Dynamic Programming. These classes are used to solve the Hamilton-Jacobi (HJ) or Hamilton-Jacobi-Bellman (HJB) equations for different optimization problems for single and multi-agent systems~\cite{Lewis_2012,Abouheaf2019_Rob,Abouheaf2017a}. This is done in order to optimize a utility function associated to the dynamic system under consideration. Devising solutions for multi-agent optimization problems can be computationally challenging due to the coupling effects in the optimality equations~\cite{AbouheafCTT2015}. Furthermore, the complexity increases even further when trying to solve that problem in a distributed fashion using only local information available to each individual agent.

Reinforcement Learning (RL) is a ML mechanism that is  adaptive to the different ADP solutions~\cite{Sutton_1998}. It is concerned with selecting the best strategy-to-apply in a dynamic learning process that penalizes or rewards such strategy using a given cost function. This enables the system to transit from one state to another which represents a better choice in the path of finding an optimal solution to the problem in hand~\cite{IMRLGhasem}. The RL process mimics the analytical solution environment of the optimization problems where the optimal strategy and its value are interrelated~\cite{Lewis_2012}. The RL platform employs temporal difference solutions arising from the various ADP classes. However, these solutions may get overwhelmingly complex especially in the case of multi-agent systems. Hence, solvers that are based on combined ideas of game theory, cooperative optimization, and RL are used to provide distributed and real-time solutions for such problems. The RL solutions are implemented using one of two two-step techniques known as value iteration and policy iteration methods. They differ at how the strategy is evaluated and then updated accordingly~\cite{Sutton_1998}. Within the same context, the adaptive actor-critic structures are approximation tools used by the RL algorithms. The actor is a neural network that assesses the quality of the adopted policy, while the critic is a neural network that estimates the cost-to-go from each state to a final following a certain policy~\cite{Abouheaf20TIM}. 



\section*{Pursuer-Evader Problem Setup}
\label{problem_fromulation}

\begin{figure}[hbt]
  \centering
  \tikzset{every picture/.style={line width=0.75pt}} 
  \begin{tikzpicture}[x=0.75pt,y=0.75pt,yscale=-1,xscale=1,scale=0.875]

\draw    (313,216) -- (465.5,215.01) ;
\draw [shift={(467.5,215)}, rotate = 539.63] [color={rgb, 255:red, 0; green, 0; blue, 0 }  ][line width=0.75]    (10.93,-3.29) .. controls (6.95,-1.4) and (3.31,-0.3) .. (0,0) .. controls (3.31,0.3) and (6.95,1.4) .. (10.93,3.29)   ;
\draw    (313,216) -- (313.99,48) ;
\draw [shift={(314,46)}, rotate = 450.34] [color={rgb, 255:red, 0; green, 0; blue, 0 }  ][line width=0.75]    (10.93,-3.29) .. controls (6.95,-1.4) and (3.31,-0.3) .. (0,0) .. controls (3.31,0.3) and (6.95,1.4) .. (10.93,3.29)   ;
\draw    (313,216) -- (278.22,261.41) ;
\draw [shift={(277,263)}, rotate = 307.45] [color={rgb, 255:red, 0; green, 0; blue, 0 }  ][line width=0.75]    (10.93,-3.29) .. controls (6.95,-1.4) and (3.31,-0.3) .. (0,0) .. controls (3.31,0.3) and (6.95,1.4) .. (10.93,3.29)   ;
\draw   (203.63,216) .. controls (203.63,155.59) and (252.59,106.63) .. (313,106.63) .. controls (373.41,106.63) and (422.38,155.59) .. (422.38,216) .. controls (422.38,276.41) and (373.41,325.38) .. (313,325.38) .. controls (252.59,325.38) and (203.63,276.41) .. (203.63,216) -- cycle ;
\draw  [color={rgb, 255:red, 0; green, 0; blue, 0 }  ,draw opacity=1 ][fill={rgb, 255:red, 184; green, 133; blue, 234 }  ,fill opacity=0.39 ] (340,143) -- (365.35,143) -- (382,180) -- (356.65,180) -- cycle ;
\draw    (360.5,162.24) -- (400,161.06) ;
\draw [shift={(402,161)}, rotate = 538.3] [color={rgb, 255:red, 0; green, 0; blue, 0 }  ][line width=0.75]    (10.93,-3.29) .. controls (6.95,-1.4) and (3.31,-0.3) .. (0,0) .. controls (3.31,0.3) and (6.95,1.4) .. (10.93,3.29)   ;
\draw    (360.5,162.24) -- (343.69,116.88) ;
\draw [shift={(343,115)}, rotate = 429.66999999999996] [color={rgb, 255:red, 0; green, 0; blue, 0 }  ][line width=0.75]    (10.93,-3.29) .. controls (6.95,-1.4) and (3.31,-0.3) .. (0,0) .. controls (3.31,0.3) and (6.95,1.4) .. (10.93,3.29)   ;
\draw    (360.5,162.24) -- (384.81,129.6) ;
\draw [shift={(386,128)}, rotate = 486.68] [color={rgb, 255:red, 0; green, 0; blue, 0 }  ][line width=0.75]    (10.93,-3.29) .. controls (6.95,-1.4) and (3.31,-0.3) .. (0,0) .. controls (3.31,0.3) and (6.95,1.4) .. (10.93,3.29)   ;
\draw [color={rgb, 255:red, 184; green, 233; blue, 134 }  ,draw opacity=1 ]   (313,216) -- (361,161.5) ;
\draw [color={rgb, 255:red, 184; green, 233; blue, 134 }  ,draw opacity=1 ]   (313,216) -- (378,254) ;
\draw    (203.63,216) .. controls (209,232) and (329,327) .. (422.38,216) ;
\draw [color={rgb, 255:red, 245; green, 166; blue, 35 }  ,draw opacity=1 ]   (313,106.63) .. controls (395.5,172) and (376.5,254) .. (378,254) ;
\draw    (299.5,233.75) .. controls (315.74,240.91) and (322.39,235.31) .. (333.41,228.91) ;
\draw [shift={(335,228)}, rotate = 510.64] [color={rgb, 255:red, 0; green, 0; blue, 0 }  ][line width=0.75]    (10.93,-3.29) .. controls (6.95,-1.4) and (3.31,-0.3) .. (0,0) .. controls (3.31,0.3) and (6.95,1.4) .. (10.93,3.29)   ;
\draw [color={rgb, 255:red, 126; green, 211; blue, 33 }  ,draw opacity=1 ]   (335,228) .. controls (351.63,208.53) and (340.69,197.68) .. (336.35,193.59) ;
\draw [shift={(335,192.25)}, rotate = 411.34000000000003] [color={rgb, 255:red, 126; green, 211; blue, 33 }  ,draw opacity=1 ][line width=0.75]    (10.93,-3.29) .. controls (6.95,-1.4) and (3.31,-0.3) .. (0,0) .. controls (3.31,0.3) and (6.95,1.4) .. (10.93,3.29)   ;

\draw (343,102.5) node  [font=\small] [align=left] {N\textsubscript{j}};
\draw (389,120) node  [font=\small] [align=left] {U\textsubscript{j}};
\draw (416,157) node  [font=\small] [align=left] {E\textsubscript{j}};
\draw (350.5,199.5) node  [font=\small]  {$\gamma $};
\draw (316,245) node  [font=\small]  {$\theta $};
\draw (273,272) node   [align=left] {X};
\draw (469,206) node   [align=left] {Y};
\draw (324,50) node   [align=left] {Z};
\draw (336.42,177.5) node  [font=\small] [align=left] {R};
%
\end{tikzpicture}
  \caption{Earth and vessel coordinate systems}
  \label{fig:Trans}
\end{figure}
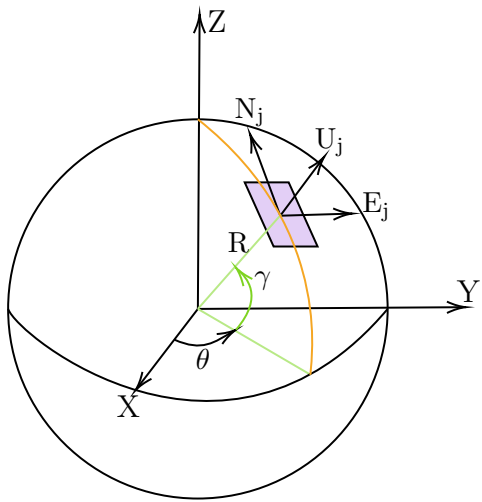

Consider a system of one evader and $N$ pursuers. Each vessel $j$ is associated with an (ENU) Cartesian coordinate system $\text{RF}_j$ pinned at the vessel's position, such that vector $\rotmatvar{E}_j$ points to the vessel's heading and the 2D plane ($\rotmatvar{E}_j$, $\rotmatvar{N}[][][][j]$) is tangent to the earth surface, which is modeled as a sphere of radius $R$ (see \Cref{fig:Trans}). Let $\rotmatvar{(\cdot)}[][i]$ denote the quantity $(\cdot)$ expressed in $\text{RF}_i$; and $({}_{}^{i}x_k^j, {}_{}^{i}y_k^j) \in \mathbb{R}^2$, ${}_{}^{i}v_k^j \in \mathbb{R}$, and ${}_{}^{i}\theta_k^j \in SO(2)$, be the x-y coordinates, linear surge speed (along $\rotmatvar{E}_j$), and the orientation angle, respectively, of vessel $j$ expressed in $\text{RF}_i$, where $j=o,1,2,\ldots,N$. Vessel $o$ denotes the evader, whereas vessels $1,2,\ldots,N$, denote the pursuers. The subscript $k$ and $SO(2)$ refer to the discrete-time index and the Special Orthogonal group in 2D, respectively.
The decision-making mechanism is executed in a decentralized fashion where the RL algorithm is run locally by each pursuer $j=1,2,\ldots,N$, with the goal of minimizing the magnitudes of the tracking errors $e^{\ell,j}_k=\ell_k^j-\ell^o_k$, for $\ell \in \{x,y\}$, and consequently $e^{\theta,j}_k=\theta_k^j-\theta^o_k$, as $k \rightarrow \infty$. This is done by computing two adjustment scalars $u^{x,j}_k$ and $u^{y,j}_k$ which are used to tune the surge velocity and heading of vessel $j$, such that
\begin{subequations}
  \begin{equation}
    {}_{}^{j} v_{k+1}^j = {}_{}^{j} v_{k}^j + \delta \, (u_{k}^{v,j} - u_{k-1}^{v,j})
    \label{vel-ad}
  \end{equation}
  \begin{equation}
    {}_{}^{j} \theta_{k+1}^j = {}_{}^{j} \theta_k^j + u_k^{\theta,j}
    \label{head-ad}
  \end{equation}
\end{subequations}
where $u_k^{v,j}=\sqrt{(u_k^{x,j})^2+(u_k^{y,j})^2}$, $u_k^{\theta,j} = \tan^{-1}({u_k^{x,j}}/{u_k^{y,j}}) \in [-\pi/2,+\pi/2]$~\si{\radian}, and $\delta$ is a discount factor.
Since the same RL algorithm is run by each vessel $j$ and the output signals are computed with respect to $\text{RF}_j$, the reference frame and the vessel identifiers will no longer be explicitly indicated in the rest of the manuscript.

The corrective adjustments $u_k^\ell$, $\ell \in \{x,y\}$, are set up as
\begin{equation}
 u^{\ell}_k=\boldsymbol{K}^{{\ell}}_k \, \boldsymbol{E}_k^{\ell}
 \label{strat}
\end{equation}
where $\boldsymbol{E}_k^{\ell} = \vect{ e_k^{\ell} , e_{k-1}^{\ell} , e_{k-2}^{\ell} }^T \in \mathbb{R}^{3}$ is a window of the last three recorded tracking errors between the pursuer and the evader along the $\ell$-direction, while $\boldsymbol{K}^{\ell}_k \in \mathbb{R}^{1 \times 3}$ represents the control gains which are computed in real-time by the RL algorithm.

One of the salient features of this setup is its flexibility to adapt to different problems of various complexities. For instance, for the application in hand, it was found sufficient to only include the last three tracking errors in $\boldsymbol{E}_k^{\ell}$. However, it is possible to add more error signals or remove some without having to redesign the control algorithm. The same remark is applicable for modeling the incremental difference in the surge speed and the orientation angle. In \eqref{vel-ad} and \eqref{head-ad}, such signals are shaped as $v_{k+1} - v_{k} = \delta \, (u_{k}^{v} - u_{k-1}^{v})$ and $\theta_{k+1} - \theta_k = u_k^{\theta}$, respectively. Changing them to some other forms, including $(u_{k}^{v}, u_{k-1}^{v}, u_{k-2}^{v}, \ldots)$, and $(u_{k}^{\theta}, u_{k-1}^{\theta}, \ldots )$, for example, does not require the restructure of the algorithm. This exhibits the flexibility inherited in the learning system where, using only few self-adapted control gains, a filtering behavior can be obtained during the online learning process.


\section*{Temporal Difference Solution}
\label{op_con}
For the RL framework to provide a solution for the pursuer-evader optimization problem, we need to derive the necessary temporal difference equations (i.e., Bellman equations) along with the required optimality conditions. The ultimate objective is to converge to the best policies $u^{v*}_k$ and $u^{\theta*}_k$ for each pursuer in order to intercept the evader ship. 
Hence, two convex objective functions are considered (each corresponds to the motion in one of the x-y directions)
\begin{eqnarray*}
U^{\ell}\left(\boldsymbol{E}_k^{\ell},{u}_k^{\ell}\right)=\frac{1}{2}\left(\boldsymbol{E}_k^{{\ell}^T} \boldsymbol{Q}^{\ell} \boldsymbol{E}^{\ell}_k+ {R}^{\ell}  \left({u}_k^{\ell}\right)^2\right)
\end{eqnarray*}
where $\ell \in \{x,y\}$, 
$\boldsymbol{0} < \boldsymbol{Q}^\ell \in \mathbb{R}^{3 \times 3}$ (positive definite) and $0 < {R}^\ell \in \mathbb{R}$.
The quality of the applied control strategies is assessed through a performance measure
%
%
\begin{equation}
  \label{Jxy_Opt}
  J_k^{\ell} = \sum_{i=k}^{\infty}U^{\ell}\left(\boldsymbol{E}_i^{\ell},{u}_i^{\ell}\right)
\end{equation}
Taking advantage of the policy expression~\eqref{strat}, the optimization process aims at minimizing the performance index $J_k^{\ell}$ by optimizing the gain vector $\boldsymbol{K}^{\ell}_k$. This iterative process may lead the obtained policy to practically converge towards the optimal policy ${u}^{\ell*}_k= \argmin_{u^\ell_k} \left(J_k^\ell\right)$, although it may not reach its theoretical value.

A common problem in RL is that $J_k^{\ell}$ cannot be calculated. Motivated by the structure of the cost function $U^\ell$, $J_k^{\ell}$ is approximated by a value function $V^\ell$ that is only dependent on two time-variant signals $\boldsymbol{E}_k^\ell$ and ${u}_k^\ell$.
\begin{eqnarray}
J_k^\ell \equiv V^\ell(\boldsymbol{E}_k^\ell,{u}_k^\ell)
\label{val_def}
\end{eqnarray}
From~\eqref{Jxy_Opt} and~\eqref{val_def}, the temporal difference equation is formed as
\begin{eqnarray}
V^{\ell}(\boldsymbol{E}_k^{\ell},{u}_k^{\ell})=U^{\ell}\left(\boldsymbol{E}_k^{\ell},{u}_k^{\ell}\right)+V^{\ell}(\boldsymbol{E}_{k+1}^{\ell},{u}_{k+1}^{\ell})
\label{Bell}
\end{eqnarray}
As with the cost function ${U}^\ell$, the value function is set as a quadratic function in the error vector $\boldsymbol{E}_k^\ell$ and the control signal ${u}_k^\ell$, such that
\begin{eqnarray*}
V^{\ell}(\boldsymbol{E}_k^{\ell},{u}_k^{\ell})=\displaystyle \frac{1}{2} [\boldsymbol{E}_k^{{\ell}^T} \,\,\,\,  {u}_k^{{\ell}^T}]\,\,\, \boldsymbol{H}^{{\ell}} \,\,\, 
\left[\begin{array}{l}
\boldsymbol{E}_k^{\ell}\\
{u}_k^{\ell}
\end{array}\right],
\label{valdef}
\end{eqnarray*}
where $\boldsymbol{H}^{{\ell}}$ is a symmetric positive definite matrix with the following block structure
\begin{equation*}
  \boldsymbol{H}^{\ell}
  =
  \left[
    \begin{array} {ll}
      \boldsymbol{H}_{EE}^{\ell}
      & \boldsymbol{H}_{Eu}^{\ell} \\
      \boldsymbol{H}_{uE}^{\ell}
      & \boldsymbol{H}_{uu}^{\ell}
    \end{array}
  \right]
\end{equation*}
Note that the structure of the solving value function plays an important role in how a RL algorithm decides on the best strategies~\cite{AbouheafCTT2015}. With the chosen quadratic structure, the pursuer-evader optimization problem is reduced to finding the optimal solving value functions, by optimizing  $\boldsymbol{H}^{\ell}$, and consequently computing the optimal control gain $\boldsymbol{K}^{\ell*}_k$ using Bellman optimality principles~\cite{Lewis_2012}. To this end, we can conclude that the optimal policy is
\begin{eqnarray}
  u_k^{{\ell}*}
  =
  \argmin_{u_k^{\ell}} V^{\ell}(\boldsymbol{E}_k^{\ell},{u}_k^{\ell})
  =
  - (\boldsymbol{H}_{uu}^{\ell*})^{-1}  \boldsymbol{H}_{uE}^{\ell*}
  \, \boldsymbol{E}_k^{\ell}
\label{OpPol}
\end{eqnarray}
Relations \eqref{strat} and \eqref{OpPol} yield the optimal gain
\begin{equation*}
  \boldsymbol{K}^{{\ell}*}
  =
  - (\boldsymbol{H}_{uu}^{\ell*})^{-1}  \boldsymbol{H}_{uE}^{\ell*}
\end{equation*}
Applying the optimal policy \eqref{OpPol} in \eqref{Bell} yields the following Bellman optimality equation
\begin{eqnarray}
V^{{\ell}*}(\boldsymbol{E}_k^{\ell},{u}_k^{{\ell}*})=U^{{\ell}*}\left(\boldsymbol{E}_k^{\ell},{u}_k^{{\ell}*}\right)+V^{{\ell}*}(\boldsymbol{E}_{k+1}^{\ell},{u}_{k+1}^{{\ell}*})
\label{Bell_Opt}
\end{eqnarray}

\section*{Online Reinforcement Learning Solution}
\label{RL}
In most real-world scenarios, it is impossible to find theoretical solutions for~\eqref{OpPol} and~\eqref{Bell_Opt}. To alleviate this problem, an online Value Iteration algorithm is employed herein to iteratively approximate such solutions. 


\subsection*{Online Value Iteration}
Value Iteration is a two-step technique that is used to find the solutions of many temporal difference forms, such as the Bellman optimality equations~\eqref{Bell_Opt}. In the first step, the solving value function is updated as
\begin{eqnarray*}
V^{{\ell}(r+1)}(\boldsymbol{E}_k^{\ell},{u}_k^{\ell})=U^{{\ell}(r)}\left(\boldsymbol{E}_k^{\ell},{u}_k^{\ell}\right) +V^{{\ell}(r)}(\boldsymbol{E}_{k+1}^{{\ell}},{u}_{k+1}^{\ell})
\label{Bell_val}
\end{eqnarray*}
where $r$ is the value-update index. The second step uses such update to suggest a new (improved) strategy, given by
\begin{eqnarray*}  
  u_k^{{\ell}(r+1)}
  =
  - \left[
  \qty( \boldsymbol{H}^{\ell}_{uu} )^{-1}  \boldsymbol{H}^{\ell}_{u E}
  \right]^{(r+1)}
  \,\,  \boldsymbol{E}_k^{\ell}  .
\label{OpPol_Val}
\end{eqnarray*}
The algorithm cycles through this process until convergence is achieved. 
%
%
The Value Iteration technique is generally proven to converge. It yields a bounded sequence of non-decreasing solving value functions if the initial solving value function is selected to be positive definite~\cite{AbouheafRob18}. The following development explains how the Value Iteration process approximates the solving value function $V^{\ell}$ and the optimal strategy $u^{\ell}$ using means of the adaptive actor-critics.

\subsection*{Adaptive Actor-Critics Approximation}

In this work, a critic neural network is applied to approximate the solving value function, whereas an actor neural network is designed to estimate the optimal strategy~\cite{Bertsekas1996,Prokhorov1997}.
In the following, we will use the notation $\hat{(\cdot)}$ to denote an estimate of $(\cdot)$.

The approximation of the solving value function for each pursuing vessel is motivated by the following quadratic structure in the $\ell$-direction, where $\ell \in \{x,y\}$:
\begin{eqnarray*}
  \hat V^{\ell}(\boldsymbol{E}_k^{\ell},{\hat u_k^{\ell}})
  =
  \frac{1}{2}
  \vect{ \boldsymbol{E}_k^{{\ell}^T} ,  {\hat u_k}^{{\ell}^T} }
  \,\,\, \boldsymbol{W}^{[c]{\ell}} \,\,\, 
  \left[\begin{array}{l}
          \boldsymbol{E}_k^{\ell}\\
          {\hat u_k^{\ell}}
        \end{array}\right]  , 
\end{eqnarray*}
with $\boldsymbol{W}^{[c]{\ell}}  \in \mathbb{R}^{4 \times 4}$ being the critic approximation weights matrix of the following block structure
\begin{equation*}
  \boldsymbol{W}^{[c]{\ell}}
  =
  \left[
    \begin{array} {ll}
      \boldsymbol{W}^{[c]{\ell}}_{EE}
      & \boldsymbol{W}^{[c]{\ell}}_{Eu} \\[1mm]
      \boldsymbol{W}^{[c]{\ell}}_{uE}
      & \boldsymbol{W}^{[c]{\ell}}_{uu}
    \end{array}
  \right] .
\end{equation*}
In a similar fashion, the optimized strategy is estimated by a form similar to that of~\eqref{strat} and~\eqref{OpPol},
\begin{eqnarray*}
\hat u_k^{\ell}= \boldsymbol{W}^{[a]{\ell}} \boldsymbol{E}_k^{\ell} , 
\end{eqnarray*}
where $\boldsymbol{W}^{[a]{\ell}} \in \mathbb{R}^{1 \times 3}$ holds the actor approximation weights.

Let $\varepsilon_k^{[c]{\ell}}$ and $\varepsilon_k^{[a]{\ell}}$ be critic and actor weight approximation errors, respectively. Then,
\begin{align*}
  \varepsilon_k^{[c]{\ell}}
  & =
    \frac{1}{2}
    \left[
    \hat V^{\ell}(\boldsymbol{E}_k^{\ell},{\hat u_k^{\ell}})
    - \tilde V^{\ell}_k
    \right]^2
  \\
  \varepsilon_k^{[a]{\ell}}
  & =
    \frac{1}{2}
    \left[
    \hat u_k^{\ell} - \tilde u^{\ell}_k
    \right]^2
\end{align*}
where
\begin{align*}
  \tilde V^{\ell}_k
  & =
    U^{\ell}
    \left(
    \boldsymbol{E}_k^{\ell},\hat {u}_k^{\ell}
    \right)
    +
    \hat V^{\ell}(\boldsymbol{E}_{k+1}^{\ell},\hat{u}_{k+1}^{\ell})
  \\
  \tilde u_k^{\ell}
  & =
    - \left[
    \qty( \boldsymbol{W}^{[c]\ell}_{uu} )^{-1}  \boldsymbol{W}^{[c]\ell}_{u E}
    \right]
    \boldsymbol{E}_k^{\ell}
\end{align*}
The update of the critic and actor weights is accomplished in real-time through a gradient descent approach, such that
\begin{align}
  \label{NNval1}
  \small
  \boldsymbol{W}^{[c]{\ell}(r+1)}
  & =
    \boldsymbol{W}^{{\ell}[c](r)}
    -
    \alpha_c \,
    \varepsilon_k^{[c]{\ell}}
    \left[\begin{array}{l}
            \boldsymbol{E}_k^{\ell}\\
            {\hat u}_k^{\ell}
          \end{array}\right]
  \vect{ \boldsymbol{E}_k^{{\ell}^T} , {\hat u}_k^{{\ell}^T} }
  \\
  \label{NNPol1}
  \boldsymbol{W}^{[a]{\ell}(r+1)}
  & =
    \boldsymbol{W}^{[a]{\ell}(r)}
    -
    \alpha_a
    \varepsilon_k^{[a]{\ell}} \boldsymbol{E}_k^{\ell}
\end{align}
where $0<\alpha_c,\alpha_a<1$ are learning rate parameters and $r$ is the index corresponding to the strategy update loop.
The adaptive actor-critic implementation of the online Value Iteration process is detailed in Algorithm~\ref{alg:alg1}. The matrix magnitude used in the algorithm is defined as the maximum absolute value of the matrix elements. 
It is important to notice that the designed RL algorithm does not use any information about the dynamics of the pursuer or evader vessels. The strategies followed by each pursuer only depend on its own location measurements and those of the evader ship.

\begin{algorithm}[hbt]
	\setstretch{1} 
	\caption{Adaptive Actor-Critic Implementation}\label{alg:alg1}
	\begin{algorithmic}[1] 
		\Require
                \Statex Positive definite $\boldsymbol{Q}^{\ell}$ and ${R}^{\ell}$, $\ell \in \{x,y\}$
		\Statex Initial tracking error vectors $\boldsymbol{E}_k^{\ell}$ and strategies $\hat u^\ell_k$
		\Statex Convergence error threshold $\Delta$
                \Statex Learning parameters $\alpha_a$ and $\alpha_c$
		\Statex Width $L$ of a moving time window 
		\Statex Maximum number of learning iterations $N_t$
		\Ensure
		\Statex Tuned neural network weights $\boldsymbol{W}^{[c]{\ell}(*)}$ and $\boldsymbol{W}^{[a]{\ell}(*)}$ 
		\Statex
                \State stop $\gets$ false
                \While {stop $=$ false}
                \State $r \gets 0$ 
                \State $\boldsymbol{W}^{[a]{\ell}(r+1)}$, $\boldsymbol{W}^{[c]{\ell}(r+1)}$ $\gets$ random values
                \State actor\_converged $\gets$ false
                \State critic\_converged $\gets$ false
                \While {$r < N_t$\AlgAnd stop $=$ false}
                \State $r \gets r+1$
		\State Calculate $U^{\ell}\left(\boldsymbol{E}_k^{\ell(r)},{u}_k^{\ell(r)}\right)$
		\State Compute $\hat V^{\ell}(\boldsymbol{E}_{k+1}^{\ell(r)},{\hat u_{k+1}^{\ell(r)}})$ and $\hat {u}_{k+1}^{\ell(r)}$ 
                \State Calculate $\varepsilon_k^{[c]{\ell}(r)}$ and $\varepsilon_k^{[a]{\ell}(r)}$
		\State Update critic and actor weights\Comment{using~\eqref{NNval1},~\eqref{NNPol1}}
                \If{$r > L$}
		\If{$\norm{\boldsymbol{W}^{[c]\ell \, (r+1-j)}-\boldsymbol{W}^{[c]\ell(r-j)}}\le \Delta$, $\forall j \in \{0,1,\ldots,L\}$,}
		\State $\boldsymbol{W}^{[c]{\ell}(*)} \gets \boldsymbol{W}^{[c]{\ell}(r+1)}$
                \State critic\_converged $\gets$ true
		\EndIf
		\If{$\norm{\boldsymbol{W}^{[a]\ell \, (r+1-j)}-\boldsymbol{W}^{[a]\ell(r-j)}}\le \Delta$, $\forall j \in \{0,1,\ldots,L\}$,}
		\State $\boldsymbol{W}^{[a]{\ell}(*)} \gets \boldsymbol{W}^{[a]{\ell}(r+1)}$
                \State actor\_converged $\gets$ true
		\EndIf
                \If{critic\_converged\AlgAnd actor\_converged}
                \State stop $\gets$ true
                \EndIf
                \EndIf
                \EndWhile
                \EndWhile
		\State \Return Tuned weights $\boldsymbol{W}^{[c]{\ell}(*)}$ and $\boldsymbol{W}^{[a]{\ell}(*)}$
	\end{algorithmic}
\end{algorithm}

\section*{Simulation Scenarios}
\label{sim}

\subsection*{Total::Perception Engine and Coordinate Transformations}
While the decision-making mechanism is realized through the RL technique listed in Algorithm~\ref{alg:alg1}, the vessels' dynamics and the evader ship behavior are simulated using Total::Perception™ (T::P), which is a state-of-the-art proprietary Intelligence, Surveillance and Reconnaissance (ISR) Systems Simulation Engine developed by Larus Technologies. It is capable of fusing large amounts of data from a multitude of sources, such as, active and passive sensing packages from GPS, Radar, and other sources. It is also able to learn complex relations among the tracked vessels to recognize any anomalous patterns in order to provide accurate information-processing strategies and continuously optimize the situational awareness mechanisms for decision support makers.
Sea vessel objects in T::P are controlled through a variety of commands, such as the vessel's reference linear speed and orientation angle. The engine also enables chosen vessels to perform different missions like patrolling and interception.
%
%
To assess the performance of the proposed solution, the heading angles and the linear speeds of the pursuer vessels are controlled by the RL approach described in Algorithm~\ref{alg:alg1}, while the evader's intelligence is provided by T::P.

\begin{figure*}[htb]
  \centering
  \includegraphics[width=0.85\textwidth]{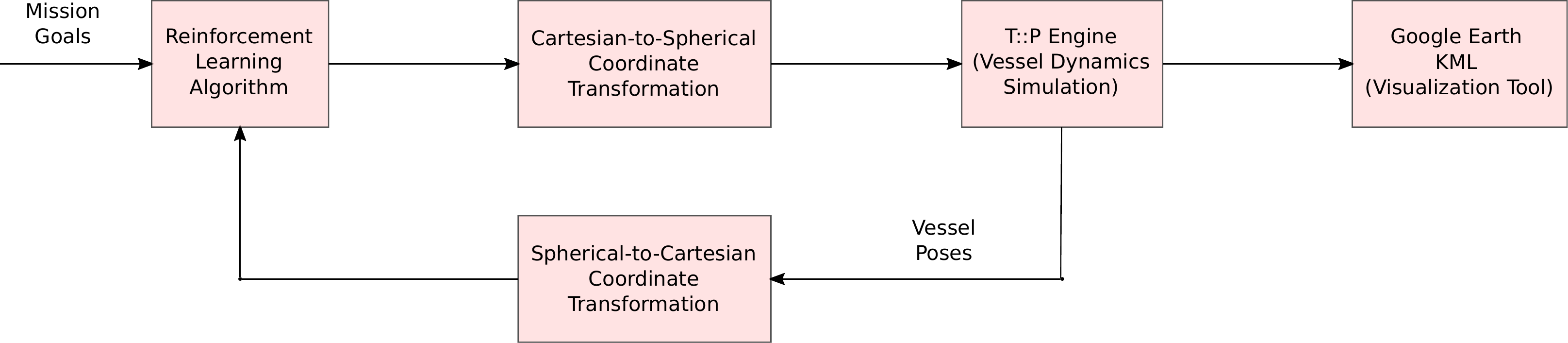}
  \caption{Simulation data flow}
  \label{fig:simulation-data-flow}
\end{figure*}

Let $\text{RF}_0$ be the (XYZ) Earth Centered Earth Fixed (ECEF) coordinates, as shown in \Cref{fig:Trans}. The T::P engine operates with spherical coordinates in $\text{RF}_0$. If the spherical coordinates of a vector \rotmatvar{u} in $\text{RF}_0$ are $( \norm{u}, \theta, \gamma )$, where $\norm{u}$ stands for the Euclidean norm, and $\theta$ and $\gamma$ are the longitude and latitude angles, respectively, then the Cartesian coordinates of \rotmatvar{u} in $\text{RF}_0$ are given by
\begin{align*}
  \rotmatvar{u}[][0]
  & =
    \norm{u} \,
    \begin{bmatrix}
	\cos(\gamma) \, \cos(\theta)\\
	\cos(\gamma) \, \sin(\theta)\\
	\sin(\gamma)
\end{bmatrix}
\end{align*}
Converting this to its equivalent Cartesian coordinates \rotmatvar{u}[][j] in the (ENU) frame $\text{RF}_j$ corresponding to vessel $j$ is performed through the rotation matrix \rotmatvar{R}[][j][][i], where $\rotmatvar{u}[][j] = \rotmatvar{R}[][j][][0] \rotmatvar{u}[][0]$ and
\begin{align*}
  \rotmatvar{R}[][j][][0]
  & =
    \begin{bmatrix}
      -\sin(\theta)  & \cos(\theta)  & 0\\
      -\cos(\theta) \, sin(\gamma)  & -\sin(\theta) \sin(\gamma)  & \cos(\gamma) \\
      \cos(\theta) \, cos(\gamma)  & \sin(\theta) \, \cos(\gamma)  & \sin(\gamma) 
	\end{bmatrix}
\end{align*}
To go back to $\text{RF}_0$, it is sufficient to multiply by the transpose of \rotmatvar{R}[][j][][0] (a property of rotation matrices). In other words, $\rotmatvar{u}[][0] = \qty( \rotmatvar{R}[][j][][0] )^T \, \rotmatvar{u}[][j]$.
For each pursuer $j$, the RL process generates the heading angle and surge speed variations or control decisions for that pursuer expressed in $\text{RF}_j$. They are transformed into their equivalent Cartesian coordinates in $\text{RF}_0$ before passing them on to the T::P engine. The updated vessel poses generated by T::P are fed back to the RL algorithm after they are transformed back to their respective vessel frame $\text{RF}_j$ coordinates.
The T::P engine interacts with Google Earth KML tool to visualize the behavior of the vessels in real-time on Google Earth~\cite{Larus}. The simulation data flow is shown in \Cref{fig:simulation-data-flow}.

\subsection*{Test Results}
To demonstrate the adaptability of the decision-making mechanism to track IUU ships, three real-world scenarios are considered.
They are realistic scenarios purposely designed by experts in the industry. They address typical situations where a suspicious IUU vessel is detected by radar and satellite monitoring systems. A resource allocation system identifies three coastguard vessels to participate in the pursuit. The command is then passed to the proposed RL-based algorithm to navigate the pursers in real-time to try to intercept the suspicious vessel as quickly as possible; preferably while it is still in action and within the Canadian marine boarders.
The simulation parameter values are listed in \Cref{tab:simulation-parameter-values}. The critic learning rate $\alpha_c$ is set to a notably small value to compensate for the excessively large initial tracking errors which are in the range of hundreds of kilometers. 
%

\begin{table}[htb]
  \renewcommand{\arraystretch}{1.3}
  \setlength{\tabcolsep}{4pt}
  \caption{Simulation parameters}
  \label{tab:simulation-parameter-values}
  \centering
  \begin{tabular}{cccccccc}
    \hline
    $\delta$ & $\boldsymbol{Q}^{\ell}$ & $R^{\ell}$ & $\alpha_a$ & $\alpha_c$ & $\Delta$ & $L$ & $N_t$\\
    \hline\hline
    $0.99$ & $10^{-4} \, I_{3\times3}$ & $0.01$ & $0.01$ & $10^{-6}$ & $10^{-8}$ & $20$ & $6,000$\\
    \hline
  \end{tabular}
\end{table}

\begin{figure}[hbt]
  \centering
  \includegraphics[width=\columnwidth]{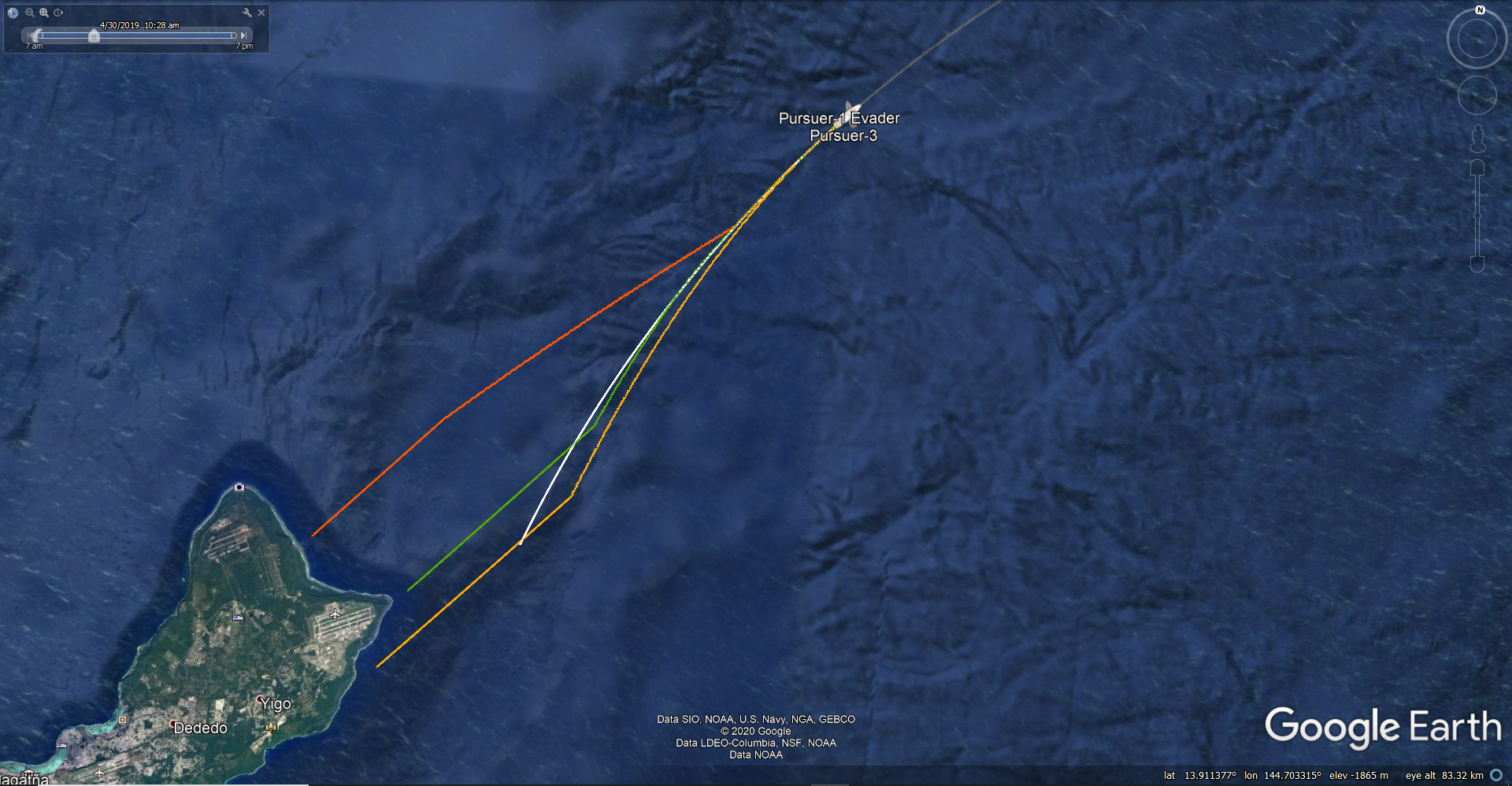}
  \caption{Intercepting the IUU ship (Scenario 1).}
  \label{fig:Sc1_Catch}
\end{figure}

In Scenario 1, the three pursers (labeled 1, 2, and 3) are given a command to intercept an IUU fishing incident. They are initially located $556$, $367$, and $\SI{289}{\km}$, respectively, apart from the IUU ship. In this scenario, the T::P engine, which is in charge of controlling the evading strategy of the evader ship, commands the ship to evade with a constant speed towards the northeast direction. \Cref{fig:Sc1_Catch} shows the trajectories of the pursuers until successfully intercepting the evader. The first vessel to reach the evader was pursuer~2, after \SI{105}{\minute}, followed by pursuers~1 and 3, which took \SI{140}{\minute} and \SI{198}{\minute}, respectively. It is interesting to notice that the first interception was not accomplished by the vessel that was initially closest to the evader. This is due to the IUU ship's evading maneuvers decided by the T::P engine.

 
  
\begin{figure}[hbt]
  \centering
  \includegraphics[width=\columnwidth]{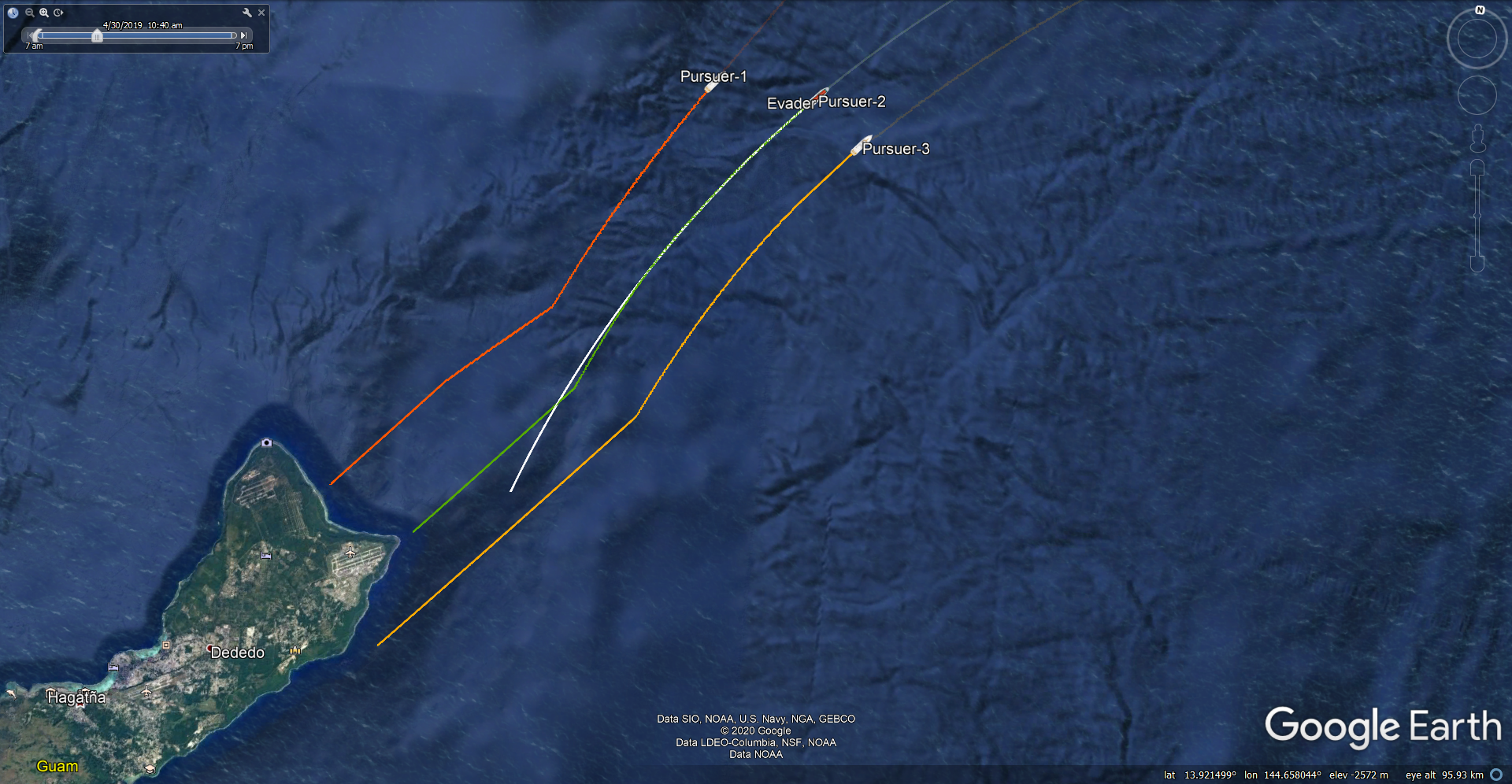}
  \caption{Intercepting the IUU ship (Scenario~2).}
  \label{fig:Sc2_Catch}
\end{figure}

In Scenario~2, the mission goals are set differently than in the previous scenario. Purser vessel~2 is now commanded to intercept the IUU ship, while vessels~1 and~3 are commanded to form a surveillance zone around the IUU ship by following it within a fixed pre-defined distance. This strategy can be useful to block the way in front of the evader to escape local waters, for instance. The vessels are initially located $744$, $378$, and $\SI{289}{\km}$ away from the IUU ship. The evader is commanded by T::P to follow the same evading strategy as in the first scenario. The outcome of the pursuit is demonstrated in \Cref{fig:Sc2_Catch}. The pursuer manage to intercept the evader in \SI{110}{\minute}, while the other two vessels draw the region which they prevent the evader to escape. This simulation shows the flexibility of the proposed RL technique in taking various forms of missions without having to modify the algorithm structure.



\begin{figure}[hbt]
  \centering
  \includegraphics[width=\columnwidth]{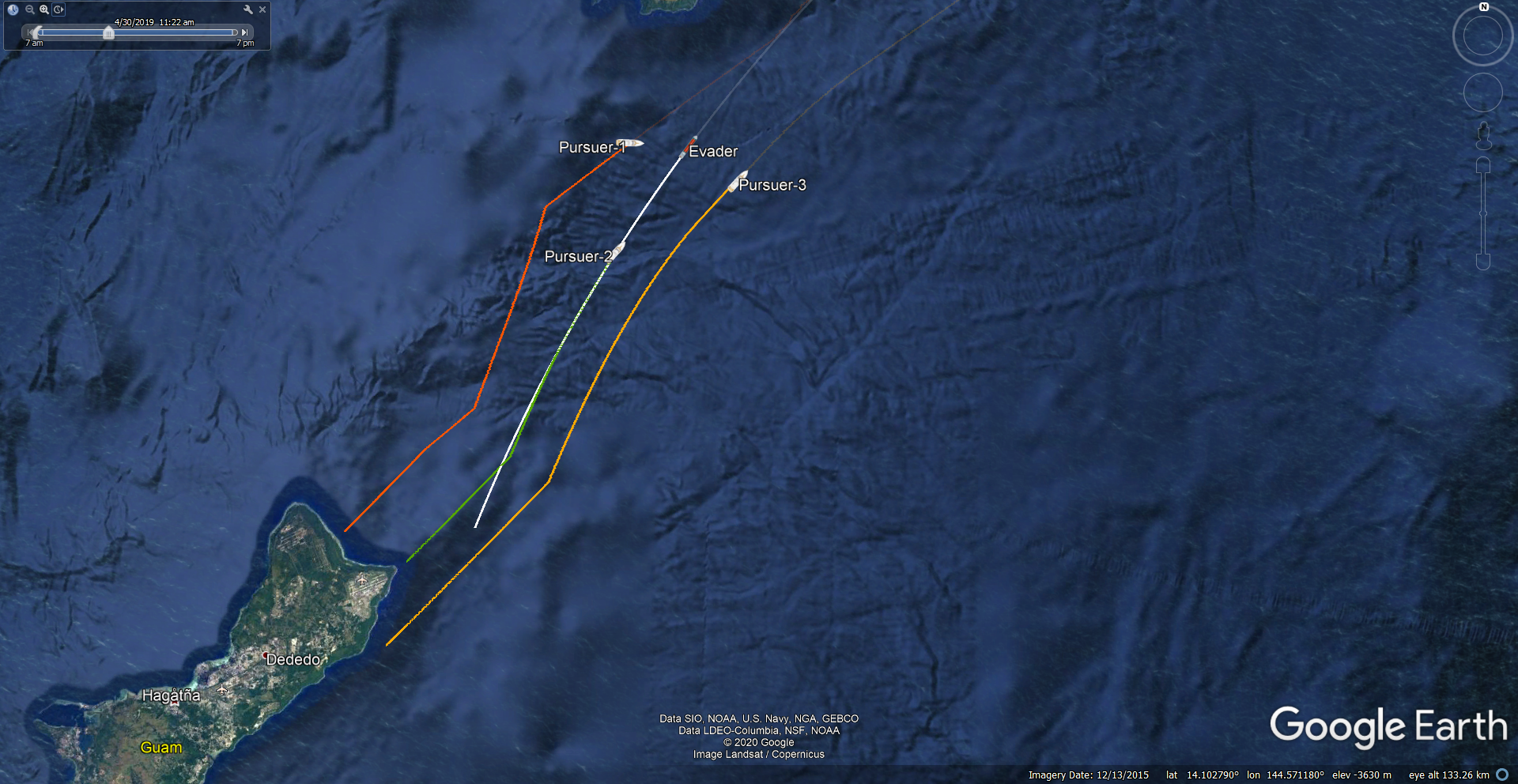}
  \caption{Vessel~2 malfunctions during the pursuit. Vessel~1 is commanded to change its mission from surveillance to interception of the evader (Scenario~3).}
  \label{fig:Sc3_Catch_malfunction}
\end{figure}

\begin{figure}[hbt]
  \centering
  \includegraphics[width=\columnwidth]{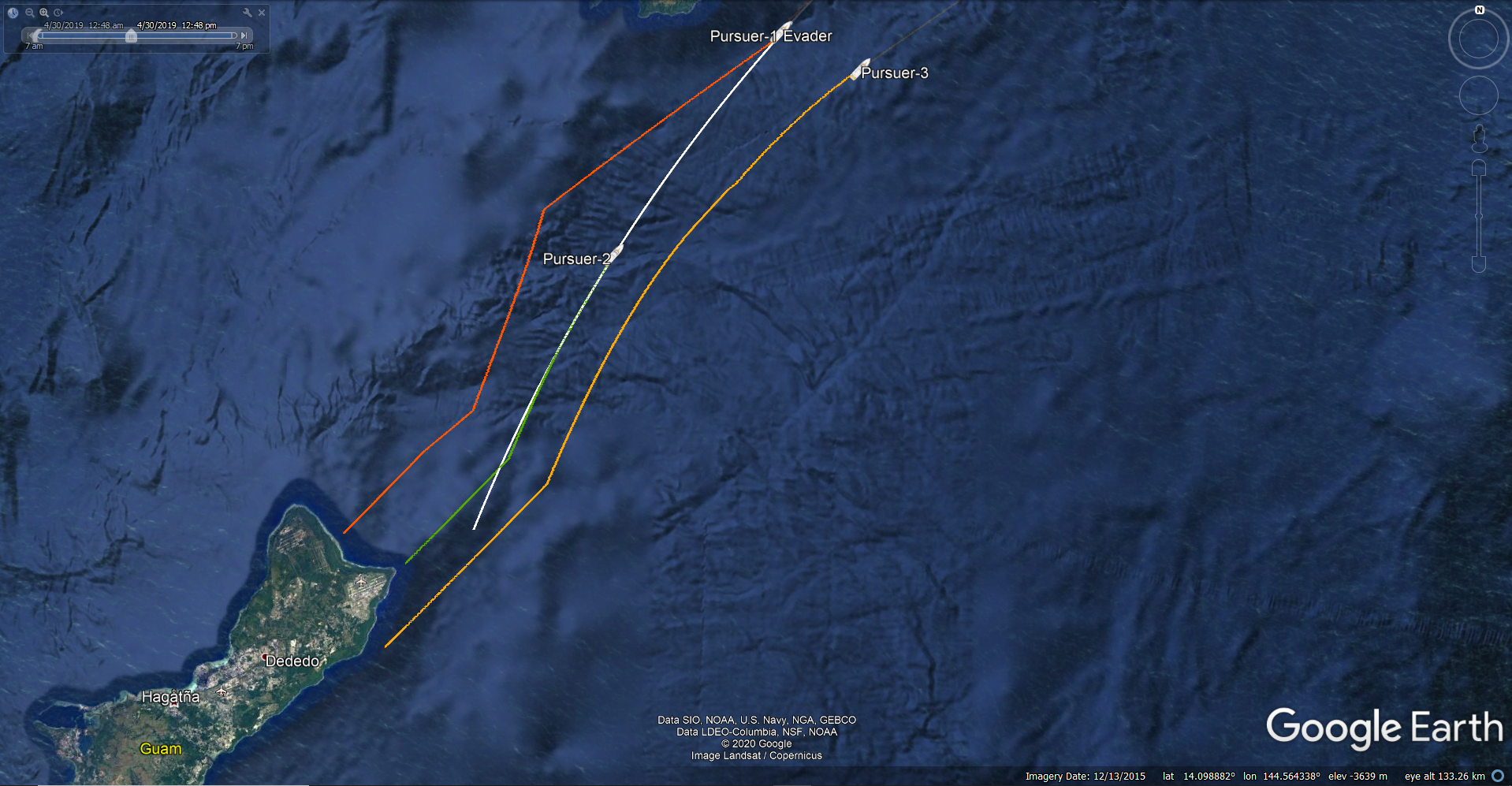}
  \caption{Pursuer vessel~1 intercepts the evader while vessel~3 surveils it from a distance (Scenario~3).}
  \label{fig:Sc3_ChangeMission}
\end{figure}

Finally, a third scenario is considered to test the robustness of the adaptive learning mechanism. To that end, Scenario~2 is repeated, except that this time, the vessel originally set to intercept the IUU ship (vessel~2) experiences a malfunction (at time \SI{117}{\minute}) after intercepting the evader in \SI{110}{\minute}, as shown in \Cref{fig:Sc3_Catch_malfunction}. At this moment, vessel~1 is commanded to change its mission from surveilling to intercepting the evader (\Cref{fig:Sc3_ChangeMission}). It took the new pursuer \SI{213}{\minute} to lock down on the evader from the moment it switched its mission from surveillance to interception. This is another example on how the proposed decision-making mechanism can easily adapt to mission changes in real-time with no extra overhead to the operator. 



\section*{Conclusion}
\label{con}
The article presented an online RL decision-making mechanism to tackle the negative impacts of IUU incidents along the Canadian coastlines. This is accomplished in real-time and in a completely decentralized fashion where the only feedback needed for each pursuing vessel is its own pose and that of the evader. No information is required about the dynamic models of the pursuing vessels or the evader ship. The adaptive learning solution is implemented using a Value Iteration process and means of neural network approximators. The solution is shown to be flexible and interactive to dynamic online variations to the pursuer missions. The proposed tool is being evaluated for the possible commercialization by Larus Technologies.



\bibliographystyle{IEEEtran}
\bibliography{Bib/mybibliography}

\begin{thebibliography}{10}
\providecommand{\url}[1]{#1}
\csname url@samestyle\endcsname
\providecommand{\newblock}{\relax}
\providecommand{\bibinfo}[2]{#2}
\providecommand{\BIBentrySTDinterwordspacing}{\spaceskip=0pt\relax}
\providecommand{\BIBentryALTinterwordstretchfactor}{4}
\providecommand{\BIBentryALTinterwordspacing}{\spaceskip=\fontdimen2\font plus
\BIBentryALTinterwordstretchfactor\fontdimen3\font minus
  \fontdimen4\font\relax}
\providecommand{\BIBforeignlanguage}[2]{{%
\expandafter\ifx\csname l@#1\endcsname\relax
\typeout{** WARNING: IEEEtran.bst: No hyphenation pattern has been}%
\typeout{** loaded for the language `#1'. Using the pattern for}%
\typeout{** the default language instead.}%
\else
\language=\csname l@#1\endcsname
\fi
#2}}
\providecommand{\BIBdecl}{\relax}
\BIBdecl

\bibitem{FAO2016}
\BIBentryALTinterwordspacing
FAO. (2016) The state of world fisheries and aquaculture: contributing to food
  security and nutrition for all. [Online]. Available:
  \url{www.fao.org/3/a-i5555e.pdf}
\BIBentrySTDinterwordspacing

\bibitem{Rami2013}
R.~Abielmona, ``Tackling big data in maritime domain awareness,''
  \emph{Vanguard}, pp. 42--43, 2013.

\bibitem{IMSANC}
P.~J.~B. {Sánchez}, M.~{Papaelias}, and F.~P.~G. {Márquez}, ``Autonomous
  underwater vehicles: Instrumentation and measurements,'' \emph{IEEE
  Instrumentation Measurement Magazine}, vol.~23, no.~2, pp. 105--114, 2020.

\bibitem{Larus}
\BIBentryALTinterwordspacing
{}{Larus Technologies}. (2020) Larus technologies products (total::insight and
  total::perception). [Online]. Available:
  \url{https://www.larus.com/products/}
\BIBentrySTDinterwordspacing

\bibitem{Rami2017}
T.~{Akinbulire}, H.~{Schwartz}, R.~{Falcon}, and R.~{Abielmona}, ``A
  reinforcement learning approach to tackle illegal, unreported and unregulated
  fishing,'' in \emph{2017 IEEE Symposium Series on Computational Intelligence
  (SSCI)}, 2017, pp. 1--8.

\bibitem{IMFogel}
D.~B. {Fogel}, ``Machine intelligence,'' \emph{IEEE Instrumentation Measurement
  Magazine}, vol.~9, no.~3, pp. 12--16, 2006.

\bibitem{IMSun}
S.~{Sun}, Y.~{Yin}, X.~{Wang}, and D.~{Xu}, ``Robust visual detection and
  tracking strategies for autonomous aerial refueling of uavs,'' \emph{IEEE
  Transactions on Instrumentation and Measurement}, vol.~68, no.~12, pp.
  4640--4652, 2019.

\bibitem{IMRLGomez}
M.~{Gomez Plaza}, T.~{Martinez-Marin}, S.~{Sanchez Prieto}, and D.~{Meziat
  Luna}, ``Integration of cell-mapping and reinforcement-learning techniques
  for motion planning of car-like robots,'' \emph{IEEE Transactions on
  Instrumentation and Measurement}, vol.~58, no.~9, pp. 3094--3103, 2009.

\bibitem{IMChap}
P.~{Chaphekarande} and D.~{Deshpande}, ``Machine learning based brain mri
  estimation method,'' in \emph{2019 2nd International Conference on
  Intelligent Computing, Instrumentation and Control Technologies (ICICICT)},
  vol.~1, 2019, pp. 1423--1430.

\bibitem{IMMoh}
S.~A. {Mohammed}, S.~{Shirmohammadi}, and S.~{Altamimi}, ``A multimodal deep
  learning-based distributed network latency measurement system,'' \emph{IEEE
  Transactions on Instrumentation and Measurement}, vol.~69, no.~5, pp.
  2487--2494, 2020.

\bibitem{Lewis_2012}
F.~Lewis, D.~Vrabie, and V.~Syrmos, \emph{Optimal Control}, 3rd~ed.\hskip 1em
  plus 0.5em minus 0.4em\relax New York, USA: John Wiley, 2012.

\bibitem{Abouheaf2019_Rob}
\BIBentryALTinterwordspacing
M.~Abouheaf, W.~Gueaieb, and D.~Spinello, ``Online multi-objective
  model-independent adaptive tracking mechanism for dynamical systems,''
  \emph{Robotics}, vol.~8, no.~4, p.~82, Sep 2019. [Online]. Available:
  \url{http://dx.doi.org/10.3390/robotics8040082}
\BIBentrySTDinterwordspacing

\bibitem{Abouheaf2017a}
M.~Abouheaf and W.~Gueaieb, ``Multi-agent reinforcement learning approach based
  on reduced value function approximations,'' in \emph{IEEE International
  Symposium on Robotics and Intelligent Sensors (IRIS)}, Oct. 2017, pp.
  111--116.

\bibitem{AbouheafCTT2015}
M.~Abouheaf, F.~Lewis, M.~Mahmoud, and D.~Mikulski, ``Discrete-time dynamic
  graphical games: Model-free reinforcement learning solution,'' \emph{Control
  Theory and Technology}, vol.~13, no.~1, pp. 55--69, 2015.

\bibitem{Sutton_1998}
R.~S. Sutton and A.~G. Barto, \emph{Reinforcement Learning: An Introduction},
  2nd~ed., ser. Second.\hskip 1em plus 0.5em minus 0.4em\relax Massachusetts:
  MIT Press, 1998.

\bibitem{IMRLGhasem}
R.~{GhasemAghaei}, M.~A. {Rahman}, W.~{Gueaieb}, and A.~{El Saddik}, ``Ant
  colony-based reinforcement learning algorithm for routing in wireless sensor
  networks,'' in \emph{2007 IEEE Instrumentation Measurement Technology
  Conference IMTC 2007}, 2007, pp. 1--6.

\bibitem{Abouheaf20TIM}
M.~{Abouheaf}, N.~Q. {Mailhot}, W.~{Gueaieb}, and D.~{Spinello}, ``Guidance
  mechanism for flexible-wing aircraft using measurement-interfaced
  machine-learning platform,'' \emph{IEEE Transactions on Instrumentation and
  Measurement}, vol.~69, no.~7, pp. 4637--4648, 2020.

\bibitem{AbouheafRob18}
M.~Abouheaf, W.~Gueaieb, and F.~Lewis, ``Model-free gradient-based adaptive
  learning controller for an unmanned flexible wing aircraft,''
  \emph{Robotics}, vol.~7, no.~4, p.~66, 2018.

\bibitem{Bertsekas1996}
D.~Bertsekas and J.~Tsitsiklis, \emph{Neuro-Dynamic Programming}, 1st~ed.\hskip
  1em plus 0.5em minus 0.4em\relax Massachusetts: Athena Scientific, 1996.

\bibitem{Prokhorov1997}
D.~Prokhorov and D.~Wunsch, ``Adaptive critic designs,'' \emph{IEEE
  Transactions on Neural Networks}, vol.~8, no.~5, pp. 997--1007, Sep. 1997.

\end{thebibliography}

\end{document}